\documentclass[prl,aps,twocolumn]{revtex4}
\usepackage{amsmath}
\usepackage{amsfonts}
\usepackage{amssymb}
\usepackage{graphicx}
\usepackage{color}

\input{tcilatex}

\begin{document}

\title{Power-law spin correlations in a perturbed honeycomb spin model}
\author{K. S. Tikhonov$^{1,2}$, M. V. Feigel'man$^{1,2}$ and A. Yu. Kitaev $%
^3$}
\affiliation{$^1$ L. D. Landau Institute for Theoretical Physics, Kosygin str.2, Moscow
119334, Russia}
\affiliation{$^2$ Moscow Institute of Physics and Technology, Moscow 141700, Russia}
\affiliation{$^3$ California Institute of Technology, Pasadena, CA 91125, USA}
\date{\today }

\begin{abstract}
We consider spin-$\frac{1}{2}$ model on the honeycomb lattice~\cite{Kitaev06}
in presence of a weak magnetic field $h_{\alpha }\ll 1$. Such a perturbation
destroys exact integrability of the model in terms of gapless fermions and 
\textit{static} $Z_{2}$ fluxes. We show that it results in appearance of a
long-range tail in the irreducible dynamic spin correlation function: $%
\left\langle \left\langle s^{z}(t,r)s^{z}(0,0)\right\rangle \right\rangle
\propto h_{z}^{2}f(t,r)$, where $f(t,r)\propto \lbrack \max (t,r)]^{-4}$ is
proportional to the density polarization function of fermions.
\end{abstract}

\maketitle

Quantum spin liquids, QSL's (see e.g. Refs.~\cite{PWA1,Wen,rev1,rev2})
present examples of strongly correlated quantum phases which do not develop
any kind of local order, while their specific entropy vanishes at zero
temperature. \emph{Critical}, or algebraic QSL's are characterized by spin
correlation functions that decay as some power of distance and time. In some
cases, the correlation asymptotics can be deduced from a representation of
spin operators in terms of almost-free fermions~\cite{Hermele05}. However, a
complete calculation based on a microscopic Hamiltonian has not been
demonstrated due to the lack of suitable exactly solvable models (in more
than one spatial dimension). We show in the present Letter that the
anisotropic spin-$\frac{1}{ 2}$ model on the honeycomb lattice, proposed by
one of us~\cite{Kitaev06}, can be used as a starting point for the
construction of an analytically treatable critical QSL. This result may seem
surprising since it is known~\cite{Baskaran07} that the original model~\cite%
{Kitaev06} possesses no spin correlations at the distances longer than a
single lattice bond. We will see however, that a small perturbation of the
model~\cite{Kitaev06}, e.g. weak external magnetic field, is sufficient to
"turn on" long-range spin correlations, albeit with a small overall
prefactor. Thus we disagree with the statement made in Ref.~\cite{Baskaran07}
that short-range character of spin correlations survives in the presence of
a weak magnetic field.

We consider the model defined by the Hamiltonian: 
\begin{equation}
\mathcal{H}=J\sum_{l=\left\langle ij\right\rangle }\left( \mathbf{\sigma }%
_{i}\mathbf{n}_{l}\right) \left( \mathbf{\sigma }_{j}\mathbf{n}_{l}\right)
-\sum_{i}\mathbf{h}_{i}\mathbf{\sigma }_{i}.  \label{H}
\end{equation}%
Unit vectors $\mathbf{n}_{l}$ are parallel to $x$, $y$ and $z$ axis for the
corresponding links $x$, $y$ and $z$ of the honeycomb lattice, see Fig.1. At 
$\mathbf{h}_{i}\equiv 0$ the Hamiltonian (\ref{H}) was solved exactly~\cite%
{Kitaev06} via a mapping to a free fermion Hamiltonian. In this approach,
each spin $\sigma _{i}$ is represented in terms of four Majorana operators $%
c_{i},~c_{i}^{x},~c_{i}^{y},~c_{i}^{z}$ with the following anticommutation
relations: $\bigl\{ c_{i}^{\alpha },c_{j}^{\beta }\bigr\}
=2\delta _{ij}\delta _{\alpha \beta }$, so that $\sigma _{i}^{\alpha
}=ic_{i}c_{i}^{\alpha }$. In terms of these new operators, the zero-field
Hamiltonian reads $\mathcal{H}=-iJ\sum_{\left\langle ij\right\rangle
}c_{i}u_{ij}c_{j}$ $\ $and $u_{ij}=ic_{i}^{\alpha }c_{j}^{\alpha }$ are
constants of motion: $\left[ \mathcal{H},u_{ij}\right] =0$, with $u_{ij}=\pm
1 $. The ground state $|G\rangle$ corresponds to an arbitrary choice of $%
\left\{u_{ij}\right\}$ that minimizes the energy. It is convenient to
introduce the notion of $Z_{2}$ flux, defined for each hexagon $\pi$ as a
product $\phi_{\pi}=\prod u_{ij}$ (since $u_{ij}=-u_{ji}$, we have to choose
a particular ordering in this definition: $i\in\text{even sublattice}$,\, $%
j\in\text{odd sublattice}$). The ground state of this model is a symmetrized
sum of states with different sets of integrals of motion $\left\{
u_{ij}\right\}$, corresponding to all fluxes equal to $1$. For practical
calculations of physical quantities, one does not have to implement such
symmetrization and can assume that all $u_{ij}\equiv 1$. We denote by $H$
the corresponding Majorana Hamiltonian: $H=-iJ\sum_{\left\langle
ij\right\rangle }c_{i}c_{j}$. It can be diagonalized with the use of Fourier
transformation. The spectrum of the resulting free fermions is gapless and
has two conic points. 
\begin{figure}[tbp]
\includegraphics[width=5.5cm,height=5cm]{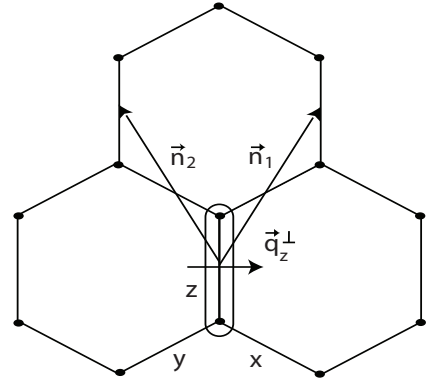}
\caption{A honeycomb lattice fragment with the $z$--$z$ link that belongs to
a given elementary cell. The indicated vector $\mathbf{q}^\perp_z = \hat{x}$
shows the direction of oscillations found in the $\langle s^z (0) s^z (%
\mathbf{{r}) \rangle }$ correlation function.}
\label{pcell}
\end{figure}
To begin, we recall the calculation of the spin-spin correlation function $%
g_{ij}^{\alpha \beta }=\left\langle \sigma _{i}^{\alpha }(t)\sigma
_{j}^{\beta }(0)\right\rangle $ in the unperturbed model with $h\equiv 0$ 
\cite{Baskaran07}. Spin operator $\sigma _{i}^{\alpha}$ acting on the ground
state produces two $Z_{2}$ fluxes and creates a fermion. Since states with
different flux configurations are mutually orthogonal and fluxes do not move
in the process of time evolution governed by the Hamiltonian $\mathcal{H}$,
a non-zero result for the correlation function is possible only if the
second spin operator $\sigma _{j}^{\beta }$ creates the same pair of fluxes.
Thus the sites $i$ and $j$ are either the same or nearest neighbors. For
larger separations between $i$ and $j$, one has $g_{ij}=0$. However, this
result is due to the static nature of $Z_{2}$ fluxes; furthermore, the
ground state is a linear combination of states with the same flux pattern.
Both these features are destroyed by any perturbation (for example, external
magnetic field) that does not commute with operators $\left\{ u_{ij}\right\}$%
.

In this paper, we consider the honeycomb lattice model with an external
magnetic field, which is treated as a weak perturbation. Before delving into
calculations, we note that a generic magnetic field opens a gap $\Delta $ in
the fermionic spectrum~\cite{Kitaev06}, with $\Delta \sim
h_{x}h_{y}h_{z}/J^{2}$. In what follows we neglect this gap. This is
definitely possible if one of the field components vanishes, i.e.\ if the
field is directed in one of the coordinate planes. For the generic field
direction, our results for spin correlations are applicable for intermediate
distances, $1\ll r\ll 1/\Delta$. For simplicity we discuss the total spin in
the $r$-th elementary cell, $s_{r}=\sigma _{r,1}+\sigma _{r,2}$ and
calculate $z$--$z$ correlations only. Since the external magnetic field
induces finite magnetization, $\left\langle s_{r}^{z}\right\rangle \neq 0$,
we study the irreducible correlation function: $g\left( t,\mathbf{r}\right)
=\left\langle \left\langle s_{r}^{z}(t)s_{0}^{z}(0)\right\rangle
\right\rangle$. We are interested in the long-time and/or long-distance
asymptotics of $g(t,\mathbf{r})$.

It is convenient to introduce complex bond fermions, defined as follows: $%
\psi _{r}=\frac{1}{2}\left( c_{r,1}+ic_{r,2}\right) $ and $\phi _{r}=\frac{1 
}{2}\left( c_{r,1}^{z}+ic_{r,2}^{z}\right) $. Operator $\phi _{r}$ creates
two fluxes in the plaquettes adjacent to the $z$-link in the elementary cell 
$r$. Note, that $\phi _{r}^{+}\phi _{r}=\frac{1+u_{ij}}{2}$, and hence the
ground-state wave function $\left\vert G\right\rangle $ satisfies $\phi
_{r}^{+}\phi _{r}\left\vert G\right\rangle =\left\vert G\right\rangle $.
Spin operator factorizes in the following way: $s_{r}^{z}=2i\psi
_{r}^{\alpha }\phi _{r}^{\alpha },$ where vectors $\psi _{r}^{\alpha}=\left(
\psi _{r},\psi _{r}^{+}\right) ,~\phi _{r}^{\alpha}=\left( \phi
_{r}^{+},\phi _{r}\right) $ are introduced and summation over $\alpha $ is
implied. In the absence of magnetic field, the correlation function of flux
operators $\phi _{r}$ is local: $G_{\phi }\left( r,t\right) =\left\langle
\phi _{r}^{+}(t)\phi _{0}(0)\right\rangle =\varphi (t)\delta _{r,0}$, which
leads to locality of the spin correlations. However, once the magnetic field
is turned on, one finds $G_{\phi }\left( r,t\right)$ to be nonzero and
proportional to $h_z^2$ at any $r$, which leads to spin correlation at large
distances. We start from the expression for $\left\langle
s_{r}^{z}(t)s_{0}^{z}(0)\right\rangle $, expanded up to the second order in $%
h_{z}$: 
\begin{eqnarray}
\left\langle s_{r}^{z}(t)s_{0}^{z}(0)\right\rangle &=&-\frac{h_{z}^{2}}{2}%
\sum_{r_{1},r_{2}}\iint d\tau _{1}d\tau _{2}\cdot  \label{g_def} \\
&&\cdot \left\langle Ts_{r}^{z}\left( t\right) s_{0}^{z}\left( 0\right)
s_{r_{1}}^{z}\left( \tau _{1}\right) s_{r_{2}}^{z}\left( \tau _{2}\right)
\right\rangle .  \notag
\end{eqnarray}%
The irreducible correlation function equals $g\left( t,\mathbf{r}\right)
=\left\langle s_{r}^{z}(t)s_{0}^{z}(0)\right\rangle -\left\langle
s_{0}^{z}\left( 0\right) \right\rangle ^{2}$, where 
\begin{equation}
\left\langle s_{0}^{z}\left( 0\right) \right\rangle =-ih_{z}\sum_{r}\int
d\tau \left\langle Ts_{0}^{z}\left( 0\right) s_{r}^{z}\left( \tau \right)
\right\rangle .
\end{equation}

Thus we have to calculate two-spin and four-spin correlation functions. For
these correlation functions to be non-zero, the flux configuration which
results from the action of the two (four) spin operators on the ground state
should coincide with the original flux configuration. In particular, the
two-spin correlator vanishes unless $r=0$, so that we have $\left\langle
Ts_{0}^{z}\left(0\right)s_{r}^{z}\left(\tau\right)\right\rangle
=u\left(\tau\right)\delta_{r,0}$ with $u\left(\tau\right)=\left\langle
Ts_{0}^{z}\left(0\right) s_{0}^{z}\left(\tau\right)\right\rangle$. It is
clear that $u(-\tau)=u(\tau)$. The expression for the magnetization thus
simplifies: 
\begin{equation}
\left\langle s^{z}\right\rangle =-2ih_{z}\int_{0}^{\infty}u\left( \tau
\right) d\tau.  \label{sz}
\end{equation}
Similarly, 
\begin{gather*}
\left\langle Ts_{r}^{z}\left( t\right) s_{0}^{z}\left( 0\right)
s_{r_{1}}^{z}\left( \tau _{1}\right) s_{r_{2}}^{z}\left( \tau _{2}\right)
\right\rangle = \\
=f_{1}\left( r,t,\tau _{1},\tau _{2}\right) \delta _{r_{1},r}\delta
_{r_{2},0}+f_{2}\left( r,t,\tau _{1},\tau _{2}\right) \delta
_{r_{1},0}\delta _{r_{2},r},
\end{gather*}
where 
\begin{gather}
f_{1}(r,t,\tau _{1},\tau _{2})=\left\langle Ts_{r}^{z}\left( t\right)
s_{0}^{z}\left( 0\right) s_{r}^{z}\left( \tau _{1}\right) s_{0}^{z}\left(
\tau _{2}\right) \right\rangle ,  \label{fdef} \\
f_{2}(r,t,\tau _{1},\tau _{2})=\left\langle Ts_{r}^{z}\left( t\right)
s_{0}^{z}\left( 0\right) s_{0}^{z}\left( \tau _{1}\right) s_{r}^{z}\left(
\tau _{2}\right) \right\rangle .  \notag
\end{gather}
In the $t\to\infty$ limit, the leading contributions to $f_1$ and $f_2$ come
from the regions $\tau_1\approx t$, $\tau_2\approx 0$ and $\tau_1\approx 0$, 
$\tau_2\approx t$, respectively. We will see that the product of spin
operators at nearby times, e.g.\ $s_{r}^{z}(t)s_{r}^{z}(\tau_2)$ in the
second case, reduces to the product of two fermion operators (up to some
renormalization). It follows that the four-spin correlation function is
asymptotically proportional to the density polarization function of free
fermions.

Note that the Wick theorem is not directly applicable to spin averages
because each spin operator creates both a fermion and some flux, the latter
acting as a scattering potential for propagating fermions. To proceed with
the calculation, one has to rewrite spin operators $s^{z}$ in equations (\ref%
{fdef}) in terms of fermions $\psi$ and $\phi$, and then move the $\phi$
operators to the right, commuting them with exponential evolution factors.
To this end, we use the identities 
\begin{equation}
\phi_{r}e^{iHt}=e^{iH_{r}t}\phi_{r},\qquad
\phi_{r}^{\dag}e^{iH_{r}t}=e^{iHt}\phi_{r}^{\dag},
\end{equation}
where the Hamiltonian $H_{r}$ differs from the original Hamiltonian $H$ by
inverting sign of the $u$ variable which belongs to the $z$-link in the
elementary cell $r$: $H_{r}=H+V_{r}$, where $V_{r} = 4J\left(
\psi_{r}^{+}\psi_{r} -\frac{1}{2} \right)$. In this way, all spin
correlators can be represented as correlators of non-interacting fermions in
the presence of external time-dependent potential. The calculation of $%
u(\tau)$ is a simple task discussed in Refs.~\cite{Baskaran07,TF10}. Using
the identity $e^{iHt}e^{-iH_{r}t}=T\exp \left( -i\int_{0}^{t}V_{r}(\tau)\,
d\tau \right)$, one arrives at the following result: 
\begin{equation*}
u(\tau) =4\left\langle T\psi _{0}\left(\tau\right) \psi_{0}^{+}\left(
0\right) e^{-i\int_{0}^{\tau}V_{0}\left(\tau^{\prime}\right) d\tau ^{\prime
}}\right\rangle\quad \text{for } \tau>0.
\end{equation*}

The next step is to calculate $f_{1,2}(r,t,\tau _{1},\tau _{2})$. We
consider explicitly all different time orderings in the expression (\ref%
{fdef}); it is enough to choose $t>0$, since $g_{r}\left( -t\right)
=g_{-r}^{\ast }\left( t\right) $: 
\begin{align*}
1)\quad & \tau _{2}>\tau _{1}>t>0; & 2)\quad & \tau _{2}>t>0>\tau _{1}; \\
3)\quad & t>0>\tau _{2}>\tau _{1}; & 4)\quad & t>\tau _{2}>\tau _{1}>0; \\
5)\quad & t>\tau _{2}>0>\tau _{1}; & 6)\quad & \tau _{2}>t>\tau _{1}>0,
\end{align*}%
while other $6$ domains $1^{\prime }..6^{\prime }$ can be obtained by the
permutation $\tau _{1}\longleftrightarrow \tau _{2}$.

Let us illustrate how to perform the calculation of $f_{1,2}^{(j)}$ for the
particular time domain $j=2$. We get for $f_{1}^{(2)}$ the following
expression (summation over $\alpha ..\delta $ is implied): 
\begin{widetext}
\begin{eqnarray}
\nonumber f_{1}^{(2)}&=&
\left\langle s_{0}^{z}\left(\tau_{2}\right) s_{r}^{z}\left(t\right)
s_{0}^{z}\left(0\right) s_{r}^{z}\left(\tau_{1}\right) \right\rangle
=16\left\langle e^{iH\tau _{2}}\psi _{0}^{\alpha }\phi _{0}^{\alpha }
e^{-iH\tau _{2}}e^{iHt}\psi _{r}^{\beta }\phi _{r}^{\beta }e^{-iHt}
\psi _{0}^{\gamma }\phi_{0}^{\gamma }e^{iH\tau _{1}}
\psi _{r}^{\delta }\phi _{r}^{\delta }e^{-iH\tau _{1}}\right\rangle =\\
\nonumber &=& 16\left\langle e^{iH\tau _{2}}\psi_{0} e^{-iH_{0}\tau_{2}}
e^{iH_{0}t} \psi_{r} e^{-iH_{r,0}t} \psi_{0}^{\dag}
e^{iH_{r}\tau _{1}} \psi_{r}^{\dag} e^{-iH\tau _{1}}
\phi_{0}^{\dag} \phi_{r}^{\dag} \phi_{0} \phi_{r}\right\rangle =\\
\label{f1} &=& -16\left\langle T\psi _{0}\left( \tau _{2}\right) \psi
_{r}\left( t\right) \psi _{0}^{+}\left( 0\right) \psi _{r}^{+}\left( \tau
_{1}\right) e^{-i\int V_{1}^{(2)}\left( \tau \right) d\tau }\right\rangle .
\\ \nonumber
\end{eqnarray}
\end{widetext}
To proceed from the first to the second line, we used the fact that the only
relevant sequence of superscripts is $\alpha \beta \gamma \delta =1122$
(recall that $\phi _{r}^{+}\phi _{r}\left\vert G\right\rangle =\left\vert
G\right\rangle ,~$while $\phi _{r}\phi _{r}^{+}\left\vert G\right\rangle =0$
and $\phi _{r}^{2}=\phi _{r}^{+2}=0$). Similarly, for $f_{2}^{(2)}$ one
obtains: 
\begin{equation}
f_{2}^{(2)}=16\left\langle T\psi _{r}\left( \tau _{2}\right) \psi
_{r}^{+}\left( t\right) \psi _{0}\left( 0\right) \psi _{0}^{+}\left( \tau
_{1}\right) e^{-i\int V_{2}^{(2)}\left( \tau \right) d\tau }\right\rangle .
\label{f2}
\end{equation}%
The potentials $V_{1,2}^{(2)}(\tau )$ are piece-wise-constant functions of
time which can be easily read off the order of fermionic operators in (\ref%
{f1}), (\ref{f2}):

\begin{center}
\begin{tabular}{|c|c|c|c|c|c|}
\hline
& $\left( -\infty ;\tau _{1}\right) $ & $\left( \tau _{1};0\right) $ & $%
\left( 0;t\right) $ & $\left( t;\tau _{2}\right) $ & $\left( \tau
_{2};\infty \right) $ \\ \hline
$V_{1}^{(2)}\left( \tau \right) $ & $0$ & $V_{r}$ & $V_{r}+V_{0}$ & $V_{0}$
& $0$ \\ \hline
$V_{2}^{(2)}\left( \tau \right) $ & $0$ & $V_{0}$ & $0$ & $V_{r}$ & $0$ \\ 
\hline
\end{tabular}
\end{center}

In the same way exact expressions for $f_{i}^{(j)},$ analogous to (\ref{f1},%
\ref{f2}), can be obtained for all other time domains $1...6$. However, they
can hardly be evaluated exactly in the closed form. The problem of their
calculation resembles the one encountered while exploring the Fermi Edge
Singularity problem~\cite{FES}, so we can analyze it similarly.

The representation of spin correlation functions in the form~(\ref{f1})
allows us to use the Wick theorem for fermions, which makes a diagrammatic
expansion of $f_{i}^{(j)}$ over the potential $V_{i}^{(j)}$ possible. Note
that apart from the normal Green function $G\left( t,\mathbf{r}\right)
=\left\langle T\psi \left( r,t\right) \psi ^{+}\left( 0,0\right)
\right\rangle $, the anomalous Green function $F\left( t,\mathbf{r}\right)
=\left\langle T\psi \left( r,t\right) \psi \left( 0,0\right) \right\rangle
=\left\langle T\psi ^{+}\left( 0,0\right) \psi ^{+}\left( r,t\right)
\right\rangle $ has also to be taken into account (we calculate both of them
below). The sum of all diagrams for each of $f_{i}^{(j)}$ is of the form $%
f_{i}^{(j)}=16e^{C_{i}^{(j)}}\cdot L_{i}^{(j)}$, where the first factor is
the the sum of closed-loop diagrams, and the second factor $L_{i}^{(j)}$ is
the a sum of open-line diagrams.

The closed-loop contribution equals $e^{C_{i}^{(j)}\left(t,\tau_{1},\tau_{2}%
\right)} =\left\langle Te^{-i\int V_{i}^{(j)}(\tau)\,d\tau}\right\rangle$.
In the limit of large time separation between pairs of points $\left\{t,\tau
_{1}\right\}$, $\left\{0,\tau_{2}\right\}$ or $\left\{t,\tau _{2}\right\}$, $%
\left\{0,\tau_{1}\right\}$, the asymptotic form of $C_{i}^{(j)}$ can be
simply determined. For example, the expressions for $C^{(2)}_{1,2}$ read: 
\begin{align*}
C_{1}^{(2)} &\approx -i\,\bigl(\Omega \left( t+\tau _{2}-\tau _{1}\right) +
\delta\Omega _{r}t\bigr), \\
C_{2}^{(2)} &\approx -i\,\Omega \left( \tau _{2}-\tau _{1}-t\right) .
\end{align*}%
In this equation, $\Omega$ is the energy of the fermionic ground state in
the presence of two adjacent fluxes ($\Omega \approx 0.04J$, see~\cite%
{Kitaev06}), while $\delta \Omega _{r}$ stands for the interaction energy of
two flux pairs separated by distance $r$. Therefore, the factor $\exp \left(
C_{i}^{(j)}\left( t,\tau _{1},\tau _{2}\right) \right) $ rapidly oscillates
with frequency $\Omega $.

Each term in the sum of open-line diagrams corresponds to a particular
pairing of four fermionic operators in the product (\ref{f1}) or (\ref{f2}).
For example, $L_{2}^{(2)}$ is given by the following equation: 
\begin{gather}
L_{2}^{(2)}=\left\langle T\psi _{r}\left( \tau _{2}\right) \psi
_{r}^{+}\left( t\right) \right\rangle _{2}^{(2)}\left\langle T\psi
_{0}\left( 0\right) \psi _{0}^{+}\left( \tau _{1}\right) \right\rangle
_{2}^{(2)}-  \label{xxx} \\
-\left\langle T\psi _{r}\left( \tau _{2}\right) \psi _{0}\left( 0\right)
\right\rangle _{2}^{(2)}\left\langle T\psi _{r}^{+}\left( t\right) \psi
_{0}^{+}\left( \tau _{1}\right) \right\rangle _{2}^{(2)}-  \notag \\
-\left\langle T\psi _{r}\left( \tau _{2}\right) \psi _{0}^{+}\left( \tau
_{1}\right) \right\rangle _{2}^{(2)}\left\langle T\psi _{0}\left( 0\right)
\psi _{r}^{+}\left( t\right) \right\rangle _{2}^{(2)},  \notag
\end{gather}
where $\left\langle T...\right\rangle _{i}^{(j)}$ stands for $\left\langle
T...e^{-i\int V_{i}^{(j)}d\tau }\right\rangle\, e^{-C_{i}^{(j)}}$.

Finally, the spin correlation function $g(t,\mathbf{r})$ is given by the
time integral over $\tau _{1},~\tau _{2}$ of the oscillating function $%
f_{i}^{(j)}$. The contribution from the domains $2$ + $2^{\prime }$ reads: 
\begin{equation}
g_{i}^{(2)}\left( t,\mathbf{r}\right) =-16h^{2}\int_{-\infty
}^{0}\int_{t}^{\infty}d\tau_{1}\,d\tau _{2}\,e^{C_{i}^{\left( 2\right)
}}L_{i}^{\left( 2\right) }.  \label{yyy}
\end{equation}
Up to this point, all calculations have been exact for any $r,t$. To proceed
further, we have to make some approximations. We use the inequality $t\gg
J^{-1}$ and average over fast oscillations of $g_{i}^{(j)}(t,\mathbf{r})$ as
a function of $t$. It is easy to see that $g_{2}^{(2)}$ has a slowly varying
part because $V_{2}^{(2)}(\tau)=0$ for $\tau\in(0,t)$. On the other hand, $%
g_{1}^{\left(2\right)}$ is purely oscillating and vanishes upon the
averaging. Considering the expression~(\ref{yyy}) for $g_{2}^{\left(
2\right)}$, we find that the main contribution to it comes from $%
\tau_{1}\approx 0$ and $\tau_{2}\approx t$ (the result of integration is
determined by a small neighborhood of the boundary points due to
oscillations of the integrand), so the corresponding expression in~(\ref{f2}%
) is of the form of fermionic density-density correlation function. Now we
have to calculate $L_{2}^{(2)}$ for $\tau_{1}\approx 0$ and $\tau
_{2}\approx t$. Note that for such time arguments, the external potential $%
V_{2}^{(2)}$ as a function of $\tau$ turns on for two short intervals (of
the order of $\Omega^{-1}$), while the separation between the pulses is
large, $t\gg J^{-1}$. In this case the long-time (or large-distance)
asymptotics of the correlation function reads $\left\langle T\psi _{r}\left(
\tau _{2}\right) \psi _{0}^{+}\left( \tau _{1}\right) \right\rangle
_{2}^{(2)}\approx G_{r}\left( t\right) \phi \left( 0-\tau _{1}\right) \bar{%
\phi}\left( \tau _{2}-t\right) ,$ where $\phi(\tau),\bar{\phi}(\tau)$ are
some dimensionless functions of $J\tau$. The double integral in~(\ref{yyy})
is thus factorized, and the renormalization due to the functions $\phi(t)$, $%
\bar{\phi}(t)$ adds an overall numerical coefficient only, which we denote
by $h_{0}^{-2}$ (it is the same for the contributions from all time
domains). Calculating the dominant first term in~(\ref{xxx}) (the other
oscillate as functions of $t$), we obtain $g_{2}^{(2)}$: 
\begin{gather*}
g_{2}^{(2)}=-16h_{z}^{2}\times \left( \int_{0}^{\infty }u\left( \tau \right)
d\tau \right) ^{2}- \\
-16h_{z}^{2}/h_{0}^{2}\left[ F\left( t,\mathbf{r}\right) F\left( -t,-\mathbf{%
r}\right) +G\left( t,\mathbf{r}\right) G\left( -t,-\mathbf{r}\right) \right]
.
\end{gather*}%
Similar considerations are applicable for the other time domains $1..6$ show
that all relevant contributions have a similar feature: the integration over 
$\left\{ \tau _{1},\tau _{2}\right\} $ is dominated by some neighborhood of
points $0$ and $t$. Collecting everything and subtracting $\left\langle
s_{0}^{z}\left( 0\right) \right\rangle ^{2},$ we obtain: 
\begin{equation}
g\left( t,\mathbf{r}\right) =-64\,\frac{h_{z}^{2}}{h_{0}^{2}}\left[ F\left(
t,\mathbf{r}\right) F\left( -t,-\mathbf{r}\right) +G\left( t,\mathbf{r}%
\right) G\left( -t,-\mathbf{r}\right) \right]  \label{result}
\end{equation}%
where free fermion Green functions $G$ and $F$ are calculated below in Eq. (%
\ref{GF1}). Thus we have found that the spin correlation function $g\left( t,%
\mathbf{r}\right)$ is proportional to the density-density correlation
function of band fermions, with the coefficient $\propto h_{z}^{2}$. The
parameter $h_{0}$ in Eq.~(\ref{result}) can be estimated (up to a numerical
constant) to be $h_{0}\sim J$ .

Now we turn to the calculation of fermionic Green functions $G\left( t,%
\mathbf{r}\right) $ and $F\left( t,\mathbf{r}\right) $ which enter (\ref%
{result}). The expression for "vector" composed of these Green functions in
the energy-coordinate representation reads: 
\begin{equation}
\left( G_{\epsilon }(\mathbf{r}),F_{\epsilon }(\mathbf{r})\right) =\frac{2i}{%
N}\sum_{\mathbf{p}}\frac{\left( (\epsilon +\Im f_{\mathbf{p}})\cos (\mathbf{%
pr}),-\Re f_{\mathbf{p}}\sin (\mathbf{pr})\right) }{\epsilon ^{2}-|f_{%
\mathbf{p}}|^{2}+i\delta \,}  \label{GF1}
\end{equation}%
where $f(\mathbf{p})=2iJ\left( 1+e^{i\mathbf{pn}_{1}}+e^{i\mathbf{pn}%
_{2}}\right) $ and $\mathbf{n}_{1,2}=\left( \pm \frac{1}{2},\frac{\sqrt{3}}{2%
}\right) $ in the standard $(x,y)$ coordinates. We expand $f(\mathbf{p})$
near the conical point $\mathbf{K}=\left( \frac{2}{3}\pi ,\frac{2}{\sqrt{3}}%
\pi \right) $ to get long-time behavior of \thinspace \thinspace\ $G\left( t,%
\mathbf{r}\right) ,$ and $F\left( t,\mathbf{r}\right) .$ Substituting these
asymptotics to Eq.(\ref{result}), we obtain the final result: 
\begin{equation}
g\left( t,\mathbf{r}\right) =\frac{16}{\pi ^{2}}\left( \frac{h_{z}}{h_{0}}%
\right) ^{2}\frac{\left( r^{2}-3\left( Jt\right) ^{2}\right) \cos ^{2}(\frac{%
2\pi }{3}\mathbf{q}_{z}^{\perp }\mathbf{r})-x^{2}}{\left( r^{2}-3\left(
Jt\right) ^{2}\right) ^{3}},  \label{final}
\end{equation}%
where $\mathbf{q}_{z}^{\perp }=\mathbf{\hat{x}}$ is the unit vector along $x$%
, orthogonal to the direction of $z-z$ link. The singularity of this
expression at $r/t=\sqrt{3}J$ is cut off by the finite width of the
Brillouin zone, which was sent to infinity while calculating integrals
leading to Eqs.(\ref{final}). The above result may seem surprising due to
the apparent anisotropy demonstrated by fast oscillations in expression~(\ref%
{final}) as a function of the $x$-component of $\mathbf{r}$. This anisotropy
is due to our choice to calculate correlations of the $z$ components of the
spin in the unit cell. Similar calculation for $x$ or $y$ components lead to
analogous results with anisotropy vectors $\mathbf{q}_{x,y}^{\perp }$, which
are perpendicular to the corresponding lattice links. The correlation
function (\ref{final}) was calculated in the lowest nontrivial order over
perturbation $h_{z}$. To account for higher-order terms in this expansion,
one should study the effects of \textit{flux motion} upon the fermion
polarization function.

In conclusions, we have shown that under weak perturbation due to magnetic
field, spin operators acquire nonzero projection $\sim h^{2}$ on the density
of band fermions, thus long-range spin correlations appear. Therefore weakly
perturbed honeycomb spin model may be considered as an example of the 
\textit{critical} QSL. We expect that the same mechanism of the coupling of
spins to fermion density can be realized for similar models on the decorated
honeycomb lattice~\cite{Yao2007,TF10}. For the gapful state~\cite{Yao2007},
we expect spin correlations to decay with a correlation length/time
determined by the gap in the fermion spectrum, whereas for the spin metal
state~\cite{TF10} asymptotic behavior $f(t,r)\propto \left[ \max (t,r)\right]
^{-2}$ is expected. Another perturbation leading to the similar effect is a
weak modification of the triad of $\mathbf{n}$ vectors, which will also
produce nonzero spin correlations at arbitrary distances.

We are grateful to L. B. Ioffe and A. S. Ioselevich for useful discussions.
This research was supported by the RFBR grant \# 10-02-00554 . 

\end{document}